\documentclass{article}
\PassOptionsToPackage{table}{xcolor}

\usepackage[preprint]{neurips_data_2024}

\usepackage{amsmath}
\usepackage[utf8]{inputenc} 
\usepackage[T1]{fontenc}    
\usepackage{hyperref}       
\usepackage{url}            
\usepackage{booktabs}       
\usepackage{amsfonts}       
\usepackage{nicefrac}       
\usepackage{microtype}      
\usepackage{xcolor}         
\usepackage{etoolbox}
\usepackage[english]{babel}
\usepackage{array}
\usepackage{tcolorbox}
\usepackage{enumitem}
\usepackage{graphicx}
\usepackage{epstopdf}
\usepackage{subcaption}
\usepackage{multirow}
\usepackage{cellspace}
\usepackage{natbib}
\setcitestyle{numbers}
\setcitestyle{comma}
\setcitestyle{square}
\setcitestyle{sort}

\DeclareMathOperator*{\argmin}{arg\,min}

\AtBeginEnvironment{tcolorbox}{\small}
\definecolor{highlight}{gray}{0.9}

\newcolumntype{C}[1]{>{\centering\arraybackslash}m{#1}}
\newcolumntype{L}[1]{>{\arraybackslash}m{#1}}

\setlist[itemize]{itemsep=0pt}

\newlength{\dataprunesubfigwidth}
\newlength{\finetunesubfigwidth}
\definecolor{efc}{rgb}{0.87, 0.19, 0.3}

\usepackage{listings}

\definecolor{codegreen}{rgb}{0,0.6,0}
\definecolor{codegray}{rgb}{0.5,0.5,0.5}
\definecolor{codepurple}{rgb}{0.58,0,0.82}
\definecolor{backcolour}{rgb}{0.95,0.95,0.92}

\colorlet{pale1}{blue!10}
\colorlet{pale2}{green!10}
\colorlet{pale3}{red!10}
\colorlet{pale4}{orange!10}
\colorlet{pale5}{cyan!10}
\colorlet{pale6}{magenta!10}
\colorlet{pale7}{gray!10}
\colorlet{pale8}{teal!10}
\colorlet{pale9}{purple!10}

\lstdefinestyle{mystyle}{
    backgroundcolor=\color{backcolour},   
    language=Python,
    basicstyle=\ttfamily\small,
    keywordstyle=\color{blue},
    commentstyle=\color{gray},
    stringstyle=\color{red},
    numbers=left,
    numberstyle=\tiny,
    numbersep=10pt
}

\title{A Language Model-Guided Framework for Mining Time Series with Distributional Shifts}

\author{
    Haibei Zhu, Yousef El-Laham, Elizabeth Fons, Svitlana Vyetrenko \\
    J.P. Morgan AI Research \\
    \texttt{\{haibei.zhu, yousef.el-laham, elizabeth.fons,} \\
    \texttt{svitlana.s.vyetrenko\}@jpmchase.com} \\
}

\begin{document}

\maketitle

\begin{abstract}

Effective utilization of time series data is often constrained by the scarcity of data quantity that reflects complex dynamics, especially under the condition of distributional shifts. Existing datasets may not encompass the full range of statistical properties required for robust and comprehensive analysis. And privacy concerns can further limit their accessibility in domains such as finance and healthcare. This paper presents an approach that utilizes large language models and data source interfaces to explore and collect time series datasets. While obtained from external sources, the collected data share critical statistical properties with primary time series datasets, making it possible to model and adapt to various scenarios. This method enlarges the data quantity when the original data is limited or lacks essential properties. It suggests that collected datasets can effectively supplement existing datasets, especially involving changes in data distribution. We demonstrate the effectiveness of the collected datasets through practical examples and show how time series forecasting foundation models fine-tuned on these datasets achieve comparable performance to those models without fine-tuning.

\end{abstract}

\section{Introduction}
\label{ch_introduction}

Time series analysis is crucial across numerous domains, including healthcare, financial, environmental science, and supply chain management, among others. Both recent advanced machine learning models and traditional statistical analyses rely heavily on the availability of comprehensive datasets that capture the underlying dynamics of the systems being studied. However, the scarcity of high-quality time series data, especially those reflecting distributional shifts, brings significant challenges, including overfitting and poor generalization of models to unseen data, to researchers. Furthermore, privacy concerns and data accessibility issues can restrict the availability of real-world datasets. Data distributional shifts, often caused by events such as pandemics, climate change, or policy changes, can significantly alter the statistical properties of the data. These shifts make it difficult for models to generalize and perform, exacerbating the issues of data scarcity.

One emerging solution to these challenges is the exploration and utilization of alternative time series datasets. These datasets are expected to share essential statistical properties with the original time series data, such as correlation, trend, and volatility. These shared properties make alternative datasets valuable proxies or supplements to existing datasets for data-hungry downstream models.

This paper proposes an approach that leverages large language models (LLMs) and data source application programming interfaces (APIs) to explore and collect time series datasets as supplements for real datasets. LLMs, such as GPT-4 and Gemini, have shown remarkable capabilities in understanding and generating text. So, we leverage the empirical knowledge provided by LLMs to optimize the data collection process through data source APIs. We believe that it is possible to collect time series data from various external data sources that share critical statistical properties with primary data and can augment the primary dataset. This approach offers several benefits:
\begin{itemize}
    \item Data augmentation: enhances the quantity and diversity of the time series datasets, providing a broader range of scenarios for downstream analysis and modeling tasks.
    \item Adaptability: collects data that meet specific requirements, such as the presence of distributional shifts in time series samples.
    \item Cost effectiveness: Leveraging publicly available data sources through APIs can reduce data acquisition costs, maintain data privacy, and ensure reproducibility.
\end{itemize}

Our contributions are as follows:
\begin{itemize}
    \item We introduce a novel framework that leverages LLMs and data APIs to efficiently collect alternative time series datasets exhibiting distributional shifts. We demonstrate the effectiveness of this approach by curating a diverse dataset collection across various domains.
    \item We showcase the utility of these datasets by fine-tuning time series forecasting foundation models and achieving comparable performance to models without fine-tuning, even in the presence of distributional shifts.
\end{itemize}

\section{Background}
\label{ch_background}

Time series data analysis has been a critical research topic in the various domains \cite{penfold2013use, tsay2005analysis, sezer2020financial, taylor2007environmental}. Traditional methods, such as autoregressive-integrated moving average (ARIMA) models \cite{box1970distribution} and exponential smoothing \cite{gardner1985exponential}, have been widely used for forecasting time series. These methods assume that the underlying data-generating process is stable over time, which often does not hold in real-world scenarios where distributional shifts occur. Recent advancements in machine learning techniques have shown promise in capturing such complex patterns in time series data with distributional shifts. However, the effectiveness of these learning-based methods is often limited by the quantity of data samples \cite{bansal2022systematic, wen2020time}. The data scarcity issue can be raised from various aspects. For example, the limited trading history of financial assets limits the quantity of financial market price time series. Regulations can also restrict access to certain datasets due to privacy and data protection concerns.

Recent studies have aimed to develop models that can effectively handle distributional shifts in time series data \cite{wang2022koopman}. These distributional shifts pose significant challenges to downstream tasks, like forecasting. Models trained on datasets without distributional shifts or datasets with limited quantity cannot capture such volatility and often struggle to adapt to these changes, leading to poor task performance. Previous studies have addressed several strategies to address the issue of data scarcity and improve model robustness against distributional shift scenarios, including creating synthetic datasets that mimic the statistical properties of the primary data \cite{yilmaz2022synthetic} and utilizing transfer learning to fine-tune models that have been trained for relevant tasks \cite{fawaz2018transfer}.

Synthetic data generation has become a significant area of research to mitigate data scarcity. Synthetic datasets can be created using various techniques, such as generative adversarial networks (GANs) \cite{yoon2019time}, which are capable of generating realistic time series data that retains the statistical properties of the original datasets. These synthetic datasets can be used to augment real data, thus enhancing the robustness of predictive models and allowing them to generalize to unseen data. Transfer learning involves adapting models trained on one domain to work effectively in another domain. This technique is useful in time series analysis, where model pre-trained on large datasets from related domains can be fine-tuned on domain-specific datasets. Domain adaptation methods, such as adversarial adaptation and discrepancy-based adaptation, can also overcome the issue of limited data quantity, aiming to align the distributions of source and target domains and making the models more robust to distributional shifts \cite{wang2018deep, shi2022deep, jin2022domain}.

Large language models, like GPT-4~\cite{openai2024gpt4} and Gemini\cite{geminiteam2024gemini}, have demonstrated the ability to understand human language and provide meaningful insights across various domains \cite{achiam2023gpt}. Beyond their linguistic capabilities, LLMs can be leveraged to identify relevant data sources and suggest optimal data retrieval strategies. In the context of time series data mining, LLMs can be leveraged to generate queries for APIs, identify potential data sources exhibiting distributional shifts, and even suggest specific periods of interest based on historical events. Data APIs provide a standardized way to access data from various online sources. Publicly available APIs, such as those offered by Google Trends, Yahoo Finance, and the Federal Reserve Economic Data (FRED), enable researchers to retrieve historical data on diverse topics. Thus, we propose this approach of leveraging LLMs and data source APIs to overcome data scarcity and effectively handle distribution shift scenarios.

\section{Dataset Mining Pipeline}
\label{ch_method}

\begin{figure}[ht]
    \caption{The overview of the dataset mining pipeline.}
    \label{fig_pipeline}
    \centering
    \includegraphics[width=\textwidth]{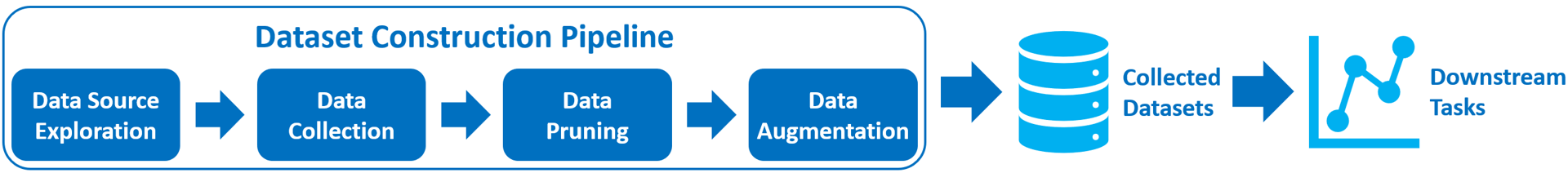}
\end{figure}

\subsection{Pipeline Overview}

The proposed method utilizes LLMs and data source APIs to collect alternative time series datasets that exhibit distributional shifts. Figure \ref{fig_pipeline} illustrates an overview of the dataset construction pipeline. This pipeline is structured into the following steps:
\begin{enumerate}
    \item Exploration and selection of data sources: leveraging the empirical knowledge provided by LLMs to identify relevant data sources that offer time series data with distributional shifts.
    \item Data collection: using queries generated from LLMs to optimize the time series data mining process through the identified data sources APIs.
    \item Data pruning: filtering the collected time series data to ensure it meets specific criteria, such as the presence of distributional shifts, using techniques like change point detection.
    \item Data augmentation: applying data transformations, such as time warping, window slicing, and window warping, to augment the quantity of collected data samples.
\end{enumerate}

\begin{figure}[ht]
    \caption{Initial prompt}
    \label{fig_init_prompt}
    \centering
    \begin{tcolorbox}[width=\textwidth]
        \textbf{Prompt to list potential data sources and APIs}
        \tcblower
        \colorbox{highlight}{Initial prompt} \\
        I want to use general-purpose LLMs such as GPT4 to assist in constructing time series datasets, with a focus on datasets that suffer from distribution shifts. Our approach does not involve training a model, just using its empirical knowledge of past events to suggest datasets and time periods that might exhibit distributional shifts. For example, S\&P500 data suffered a distribution shift during COVID-19. I want an LLM to generate query terms and data sources to build a heterogeneous time series dataset from different domains with distributional shifts. Please provide a list of open time series datasets from different contexts that can be used to query and extract time series with distribution shifts. In the list, clarify if the dataset has an API.
        Provide the list in latex tabular format with the following columns: Domain, Name of dataset, Description, API (yes/no), Link, Licence. Leave that column free if you don't have the link or license.\\
        \colorbox{highlight}{Follow up question: provide more sources} \\
        Provide additional data sources in the same format.
    \end{tcolorbox}
\end{figure}

\subsection{Exploration and Selection of Data Sources}

The initial stage of our pipeline involves the identification and selection of suitable data sources. This process leverages Large Language Models (LLMs) as extensive knowledge repositories, capitalizing on their vast training data and ability to understand natural language prompts. We craft specific prompts (Figure \ref{fig_init_prompt}) to elicit information regarding publicly available time series datasets, including their domains, descriptions, licenses, API availability, and crucially, their potential to exhibit distributional shifts due to significant events (e.g., economic crises, pandemics, policy changes). Four datasets, including Yahoo Finance, Fred Economics, EIA Energy, and Google Search Trend, have been identified by LLMs with detailed data sample information. And we focus on these datasets to collect time series samples with distributional shifts.

\begin{figure}[ht]
    \caption{Example prompt to generate queries within a certain API}
    \label{fig_prompt_query}
    \centering
    \begin{tcolorbox}[width=\textwidth]
        \textbf{Prompt to generate queries}
        \tcblower
        \colorbox{highlight}{Initial prompt: Asking for Python code to query a specific API} \\
        Let's focus on FRED. Please provide Python code to query the API to download data that might exhibit distribution shifts. I will do statistical tests to prune the data and only keep the relevant data. \\
        \colorbox{highlight}{Follow up question: Request queries} \\
        Provide a list of 50 queries for the FRED dataset in Python format with the series\_id and time ranges that I can use to download the data that exhibit distribution shifts.
    \end{tcolorbox}
\end{figure}

\subsection{Data Collection}

Upon identifying promising datasets, we proceed to the data collection stage, which extends beyond a purely technical data extraction process.  Instead, it leverages the LLM's extensive knowledge base to select time series data exhibiting a high propensity for distributional shifts.

For each identified dataset, we engage the LLM in two steps. First, we provide the LLM with the dataset's API documentation (if available) or a description of the API's structure. Importantly, we also inform the LLM of any known API limitations, such as rate limits or data access restrictions, as shown in Figure \ref{fig_prompt_query}. The LLM then generates Python code snippets tailored to interact with the specific API, taking these limitations into account. For example, the generated code may include logic for handling rate limits by pacing requests or incorporating retry mechanisms.

Second, we harness the LLM's understanding of historical events and their potential impact on time series data to construct a set of queries for each API. To facilitate programmatic interaction, these queries are structured as JSON objects, each containing a unique identifier for the specific time series within the dataset, the start and end dates for the time period of interest, and a comment justifying the selection. This justification explains why this particular time series and time period are hypothesized to exhibit distributional shifts, drawing upon the LLM's knowledge of relevant historical events such as economic recessions, natural disasters, or policy changes. Figure \ref{fig_example_query} shows examples of API-specific queries generated by the LLM.

\begin{figure}[ht]
    \caption{Example queries generated by the LLM.}
    \label{fig_example_query}
    \begin{tcolorbox}[colback=blue!5!white,colframe=white]
        \textbf{Query example for FRED API:} \\
        \texttt{\{"series\_id": "UNRATE", "start\_date": "2007-01-01", "end\_date": "2013-01-01", "comment": "Covers the Great Recession period, showcasing shifts in employment levels."\}} \\
        \\
        \textbf{Query example for EIA API:} \\
        \texttt{\{"api\_route": "electricity/rto/daily-region-data/data", "params": {"frequency": "daily", "data[0]": "value", "facets[respondent][]": "PJM", "sort[0][column]": "period", "sort[0][direction]": "desc", "offset": 0, "length": 5000, "start": "2017-09-01", "end": "2018-02-28"}, "comment": "Hurricane Maria caused significant power disruption in the PJM region."\}}
    \end{tcolorbox}
\end{figure}

This query generation process is a critical step that infuses the dataset with domain-specific knowledge and historical context. By utilizing the LLM's understanding of real-world events, we enhance the likelihood of capturing meaningful distributional shifts within the collected data. Following query generation, we perform a refinement step to eliminate any duplicates or inconsistencies. We then execute the generated Python code, utilizing the refined queries to interact with the respective APIs. This automated process efficiently and systematically downloads the identified time series data, resulting in a comprehensive collection of samples across various domains, each with a high potential for exhibiting distributional shifts.

\subsection{Data Pruning}

The collection of alternative time series datasets from various sources often results in diverse data samples with varying properties and characteristics. After acquiring the time series in the data collection step, we want to discard samples that we suspect will not be useful in our downstream use-case. For example, since our use-case of the collected time series dataset is to fine-tune a time series foundation model on distributional shift data or to construct a benchmarking dataset for evaluating the performance of forecasting models in the case of distributional shift, we want to eliminate any samples where a distribution shift is not present via a \emph{data pruning} step. The data pruning step in our pipeline takes as input all time series collected in the data collection step and outputs the subset of time series whose statistical properties satisfy a pre-defined set of requirements. In the use-case covered in this work, we require that our collected data samples exhibit distributional shifts; therefore, the data pruning step in this work utilizes \emph{offline} change point detection (CPD) as a means to pruning the time series samples. Change point detection algorithms identify points in time series samples where the statistical properties, such as mean and variance, significantly change. In this work, we utilize the \emph{ruptures} Python library \cite{truong2020selective}, which includes offline change point detection algorithms for segmenting the collected time series samples. It offers a variety of algorithms to detect change points based on different statistical metrics, such as the mean, variance, and even kernel-based approaches. By leveraging this library, we ensure the resulting data samples contain time series value distributional shifts for downstream analyses.

Let $x^{(n)}=(x_{t_1^n}^{(n)}, x_{t_2^n}^{(n)}, \ldots, x_{t_K^n}^{(n)})$ denote a time series sampled at time points ${\cal T}^{(n)}=\{t_1^n, t_2^n, \ldots, t_K^n\}$. The data pruning step categorizes each time series sample $x^{(n)}$ into one of two categories:
\begin{itemize}
\item ${\cal H}_0$: no change points detected; and
\item ${\cal H}_1$: at least one change point detected.
\end{itemize}
In the case of an unknown number of change points, a CPD scheme in the ruptures library segments the time series samples to minimize the following objective function:
\begin{equation*}
{\cal L}(x, {\cal C}) = \underbrace{\sum_{k=1}^{|{\cal C}|} \sum_{t\in{[\nu_{k-1}, \nu_k)}} V(x_t, x_{\nu_{k-1}:\nu_k})}_{\rm cost \ function} + \underbrace{\mathrm{pen}({\cal C})}_{\mathrm{penalty}},
\end{equation*}
where ${\cal C}$ denotes the set of change points, the function $V$ measures the goodness-of-fit of each time series segment, and the function $\mathrm{pen}$ is a regularizer that penalizes the objective function based on the complexity of the segmentation (e.g., for detecting changes that are too small).
If the time series can be categorized by a parametric probability model, then the cost function corresponds to the negative log-likelihood of the time series segment evaluated at the maximum likelihood estimator of the parameter. For example, in this work, the function $V$ corresponds to the Euclidean distance between each sample of the time series and the local sample mean, which is proportional to the negative log-likelihood of a Gaussian distribution evaluated at the maximum likelihood estimator of the mean (e.g., the sample mean).
Once a cost function is chosen, a search method, such as binary segmentation, is used to solve the discrete optimization problem:
\begin{equation*}
{\cal C}^\star = \argmin_{{\cal C}} {\cal L}(x, {\cal C}).
\end{equation*}
Finally, once the segmentation is established, the time series is categorized as follows:
\begin{itemize}
\item If $|{\cal C}|=1$ (i.e., only a single change point is detected at the final sample of the time series), categorize the time series as ${\cal H}_0$.
\item Otherwise, if at least one distribution shift is detected, the time series is categorized as ${\cal H}_1$.
\end{itemize}

\subsection{Data Augmentation}

Our proposed targeted approach to data collection, while efficient, may result in a smaller dataset compared to downloading entire databases. Therefore, data augmentation becomes essential to increase the quantity and diversity of our collected data, improving the robustness and generalizability of downstream models.

In this step in the pipeline, we apply three well known time series data augmentation methods with a focus on warping the time dimension to the pruned data samples: \textit{time warping}~\cite{um2017tw}, \textit{window warping}~\cite{le2016data} and \textit{window slicing}~\cite{le2016data}. These methods create new samples that retain the statistical properties of the original data while introducing variations. They primarily affect the lower frequencies of the time series data, corresponding to the trend and seasonality components, which are often more relevant for capturing the underlying dynamics and distributional shifts in the data.

Time warping alters the time intervals between data points without changing the overall shape of the time series. This simulates variations in the speed of events, making models more resilient to temporal variations. The warping path is defined by a smooth cubic spline-based curve with three knots and the knots have random magnitudes with $\mu=1$ and $\sigma=0.2$. Window warping applies random scaling to specific windows or segments within the time series, simulating localized anomalies or changes in the data distribution. This is particularly useful for modeling events like sudden market shocks or policy changes that affect specific periods. The implementation selects a random window, that is $10\%$ of the original time series length and warps the time dimension by 0.5 times or 2 times.
Window slicing extracts subsequences from the original time series by selecting random starting points within it. This method reduces the length of the time series while preserving its essential characteristics. The starting point of the window slice is chosen at random and the length of the slice is $90\%$ of the original time series. The time series is then interpolated back to the original length, therefore introducing variations while maintaining the overall structure. 

Building upon the findings of \cite{iwana2021}, we incorporated window warping and slicing into our augmentation strategy, utilizing the same hyperparameters that demonstrated consistent positive effects across diverse time series datasets and models. However, as observed in Iwana et al. (2021), time warping, when adhering to the original paper's parameters, led to performance degradation. To mitigate this, we adjusted the hyperparameters of the time warping method and verified that the augmented time series still exhibited distributional shifts.

\subsection{Collected Datasets}

\begin{table}[ht]
    \caption{Collected Datasets}
    \label{table_datasets}
    \renewcommand{\arraystretch}{1.1}
    \small
    \centering
    \resizebox{\textwidth}{!}{
    \begin{tabular}{C{14mm}C{14mm}L{40mm}C{7mm}C{7mm}C{14mm}C{14mm}C{14mm}}
        \toprule
        \multirow{2}{*}{\vspace{-5mm} \centering Name} & \multirow{2}{*}{\vspace{-5mm} \centering Domain} & \multirow{2}{*}{\vspace{-5mm} \centering Description} & \multicolumn{2}{c}{Length} & \multicolumn{3}{c}{Sample Quantity} \\
        \cmidrule(lr){4-5} \cmidrule(lr){6-8}
        & & & Min & Max & Original & After Pruning & After Augmentation \\
        \midrule
        FRED & Economics \& Finance & Macroeconomic and financial time series data & 31 & 1305 & 241 & 77 & 2310 \\
        World Cup search trends & Google search & Time series data of the popularity of World Cup 2022-related search queries on Google & 120 & 120 & 173 & 67 & 2010 \\
        EIA Daily & Energy & Time series data related to electricity generation, demand, etc. & 32 & 254 & 3750 & 1194 & 35820 \\
        Yahoo Finance & Finance & Financial market data, including stock prices, commodities, and foreign exchange & 41 & 252 & 369 & 91 & 2730 \\
        COVID search trends & Google search & Time series data of the popularity of COVID-related search queries on Google & 120 & 120 & 144 & 68 & 2040 \\
        \bottomrule
    \end{tabular}
    }
\end{table}

The resulting datasets from this pipeline include time series data from various domains, exhibiting distributional shifts in data samples. The diverse range of datasets ensures that downstream models developed on this data can generalize well to unseen data, which may have various length and temporal properties. Table \ref{table_datasets} summarizes the collected datasets, detailing their domains, descriptions, lengths, and sample quantities at each stage of the pipeline.

\section{Utility Examples}
\label{ch_experiment}

We demonstrate the effectiveness of the collected datasets by fine-tuning time series forecasting foundation models, Lag-Llama \cite{rasul2024lagllama} and Chronos \cite{ansari2024chronos}, and comparing their zero-shot prediction performance versus after fine-tuning. The experiments highlight the utility of the collected datasets in improving foundation models forecasting performance, especially in distributional shift scenarios.

\begin{table}[ht]
    \caption{Training and Testing Data Samples}
    \label{table_train_test}
    \renewcommand{\arraystretch}{1.1}
    \small
    \centering
    \begin{tabular}{C{20mm}C{35mm}C{25mm}C{25mm}}
        \toprule
        & Name & Number of fine-tuning samples & Number of testing samples \\
        \midrule
        \multirow{3}{*}{\parbox{20mm}{\vspace{-1mm}\centering In-sample\\datasets}} & FRED & 1848 & 16 \\
        & World Cup search trends & 1608 & 14 \\
        & EIA Daily & 28656 & 239 \\
        \midrule
        \multirow{2}{*}{\parbox{20mm}{\vspace{-1mm}\centering Out-sample\\datasets}} & Yahoo Finance & N/A & 91 \\
        & COVID search trends & N/A & 68 \\
        \bottomrule
    \end{tabular}
\end{table}

\subsection{Experiment Setup}

We use two recently developed time series foundation models for our experiment:
\begin{itemize}
    \item Lag-Llama: it is a foundation model designed for capturing long-term dependencies and complex temporal patterns in time series data. Lag-Llama excels in identifying and modeling lagged relationships within time series, making it effective for tasks involving sequences with extended temporal dependencies.
    \item Chronos: it is a framework for pretrained probabilistic time series models, leveraging transformer-based language model architectures. It tokenizes time series values through scaling and quantization into a fixed space. The Chronos models offer five different model sizes with parameter ranges from 20M to 710M.
\end{itemize}

The collected datasets mentioned in Section \ref{ch_method} are utilized for foundation model fine-tuning and evaluation. Specifically, the FRED, World Cup search trends, and EIA Daily energy datasets are used for fine-tuning as in-sample datasets. Once the in-sample datasets have been collected and pruned, we split them into training and testing sets following the commonly used 80-20 ratio. The training sets are augmented to let the models learn diverse patterns and scenarios present in the data, enhancing their ability to generalize and perform accurately on the testing sets. The number of time series samples for fine-tuning and testing across the three in-sample datasets and two out-sample datasets are listed in Table \ref{table_train_test}.

We evaluate the performance of the time series foundation models in two scenarios -- zero-shot prediction and prediction after fine-tuning. The zero-shot scenario evaluates the model's ability to generalize from its existing knowledge, demonstrating its capability to handle unseen datasets. The fine-tuning process will first train the existing foundation models on the collected datasets to adapt to the patterns and characteristics within the data samples. For this experiment, the collected datasets serve as the source of training data, helping the model to adapt to distributional shifts. The evaluation metrics for model prediction performance include the average mean square error (MSE), the variance of MSE across prediction samples, and the mean absolute error (MAE) coverage. The MSE measures the average squared difference between the predicted and actual time series. Lower MSE values indicate better prediction performance. The variance of MSE captures the variability of the prediction errors, showing the consistency of the prediction outcome. The MAE coverage measures the mean absolute error between the observed coverage and the target quantile levels, with lower MAE coverage values indicating more accurate and reliable prediction intervals.

\begin{table}[ht]
    \caption{Model prediction results on in-sample datasets.}
    \label{table_results}
    \small
    \centering
    \resizebox{\textwidth}{!}{
    \begin{tabular}{C{10mm}C{10mm}C{14mm}C{20mm}C{10mm}C{10mm}C{10mm}C{10mm}C{10mm}}
        \toprule
        \multirow{2}{*}{\parbox{10mm}{\vspace{5mm}\centering Model}} & \multirow{2}{*}{\parbox{10mm}{\vspace{3mm}\centering Model size}} & \multirow{2}{*}{\parbox{10mm}{\vspace{3mm}\centering Evaluation type}} & \multirow{2}{*}{\parbox{10mm}{\vspace{5mm}\centering Metrics}} & \multicolumn{3}{c}{In-sample datasets} & \multicolumn{2}{c}{Out-sample datasets} \\
        \cmidrule(lr){5-7}
        \cmidrule(lr){8-9}
        & & & & FRED & World Cup & EIA & Yahoo & COVID \\
        \midrule
        \multirow{6}{*}{\parbox{10mm}{\vspace{2mm}\centering Lag-Llama}} & \multirow{6}{*}{\parbox{10mm}{\vspace{2mm}\centering (2.5M)}}  & \multirow{3}{*}{\parbox{14mm}{\vspace{0mm}\centering Zero-shot}} & MSE & 0.1959 & 0.0126 & 0.1147 & 0.0613 & 0.0496 \\
        & & & Variance & 0.0110 & 0.0003 & 0.0082 & 0.0060 & \cellcolor{pale1} 0.0020 \\
        & & & MAE coverage & \cellcolor{pale1} 0.2646 & 0.4643 & \cellcolor{pale1} 0.3575 & 0.3346 & \cellcolor{pale1} 0.3588 \\
        \cmidrule(lr){3-4}
        & & \multirow{3}{*}{\parbox{14mm}{\vspace{0mm}\centering After\\fine-tuning}} & MSE & \cellcolor{pale1} 0.0779 & \cellcolor{pale1} 0.0105 & \cellcolor{pale1} 0.0428 & \cellcolor{pale1} 0.0488 & \cellcolor{pale1} 0.0450 \\
        & & & Variance & \cellcolor{pale1} 0.0015 & 0.0003 & \cellcolor{pale1} 0.0009 & \cellcolor{pale1} 0.0021 & 0.0032 \\
        & & & MAE coverage & 0.2910 & \cellcolor{pale1} 0.3556 & 0.3860 & \cellcolor{pale1} 0.2584 & 0.3611 \\
        \midrule
        \multirow{24}{*}{\parbox{10mm}{\vspace{12mm}\centering Chronos}} & \multirow{6}{*}{\parbox{10mm}{\vspace{2mm}\centering Tiny (8M)}} & \multirow{3}{*}{\parbox{14mm}{\vspace{0mm}\centering Zero-shot}} & MSE & 0.1403 & 0.0095 & 0.0781 & 0.0508 & 0.0514 \\
        & & & Variance & 0.0060 & 0.0002 & 0.0045 & 0.0041 & 0.0041 \\
        & & & MAE coverage & \cellcolor{pale1} 0.2330 & 0.4857 & \cellcolor{pale1} 0.2644 & \cellcolor{pale1} 0.2427 & 0.3145 \\
        \cmidrule(lr){3-4}
        & & \multirow{3}{*}{\parbox{14mm}{\vspace{0mm}\centering After\\fine-tuning}} & MSE & \cellcolor{pale1} 0.0956 & \cellcolor{pale1} 0.0054 & \cellcolor{pale1} 0.0244 & \cellcolor{pale1} 0.0445 & \cellcolor{pale1} 0.0420 \\
        & & & Variance & \cellcolor{pale1} 0.0032 & 0.0002 & \cellcolor{pale1} 0.0005 & \cellcolor{pale1} 0.0030 & \cellcolor{pale1} 0.0028 \\
        & & & MAE coverage & 0.2667 & \cellcolor{pale1} 0.4036 & 0.3582 & 0.2804 & \cellcolor{pale1} 0.2908 \\
        \cmidrule(lr){2-9}
        & \multirow{6}{*}{\parbox{10mm}{\vspace{2mm}\centering Mini (20M)}} & \multirow{3}{*}{\parbox{14mm}{\vspace{0mm}\centering Zero-shot}} & MSE & 0.1409 & 0.0108 & 0.0785 & 0.0483 & 0.0589 \\
        & & & Variance & 0.0064 & 0.0002 & 0.0049 & 0.0040 & 0.0061 \\
        & & & MAE coverage & \cellcolor{pale1} 0.2330 & 0.4857 & \cellcolor{pale1} 0.2664 & \cellcolor{pale1} 0.2425 & 0.3548 \\
        \cmidrule(lr){3-4}
        & & \multirow{3}{*}{\parbox{14mm}{\vspace{0mm}\centering After\\fine-tuning}} & MSE & \cellcolor{pale1} 0.1039 & \cellcolor{pale1} 0.0054 & \cellcolor{pale1} 0.0194 & \cellcolor{pale1} 0.0460 & \cellcolor{pale1} 0.0421 \\
        & & & Variance & \cellcolor{pale1} 0.0048 & 0.0002 & \cellcolor{pale1} 0.0004 & \cellcolor{pale1} 0.0034 & \cellcolor{pale1} 0.0028 \\
        & & & MAE coverage & 0.2719 & \cellcolor{pale1} 0.3722 & 0.3634 & 0.2825 & \cellcolor{pale1} 0.2965 \\
        \cmidrule(lr){2-9}
        & \multirow{6}{*}{\parbox{10mm}{\vspace{2mm}\centering Small (46M)}} & \multirow{3}{*}{\parbox{14mm}{\vspace{0mm}\centering Zero-shot}} & MSE & 0.1428 & 0.0113 & 0.0764 & 0.0519 & 0.0641 \\
        & & & Variance & 0.0070 & 0.0002 & 0.0043 & 0.0045 & 0.0062 \\
        & & & MAE coverage & \cellcolor{pale1} 0.2365 & 0.4893 & \cellcolor{pale1} 0.2683 & \cellcolor{pale1} 0.2324 & 0.3388 \\
        \cmidrule(lr){3-4}
        & & \multirow{3}{*}{\parbox{14mm}{\vspace{0mm}\centering After\\fine-tuning}} & MSE & \cellcolor{pale1} 0.1013 & \cellcolor{pale1} 0.0059 & \cellcolor{pale1} 0.0158 & \cellcolor{pale1} 0.0490 & \cellcolor{pale1} 0.0392 \\
        & & & Variance & \cellcolor{pale1} 0.0036 & 0.0002 & \cellcolor{pale1} 0.0004 & \cellcolor{pale1} 0.0036 & \cellcolor{pale1} 0.0028 \\
        & & & MAE coverage & 0.2858 & \cellcolor{pale1} 0.3694 & 0.3605 & 0.2712 & \cellcolor{pale1} 0.3116 \\
        \cmidrule(lr){2-9}
        & \multirow{6}{*}{\parbox{10mm}{\vspace{2mm}\centering Base (200M)}} & \multirow{3}{*}{\parbox{14mm}{\vspace{0mm}\centering Zero-shot}} & MSE & 0.1442 & 0.0173 & 0.0753 & \cellcolor{pale1} 0.0474 & 0.0681 \\
        & & & Variance & 0.0049 & 0.0003 & 0.0046 & 0.0039 & 0.0069 \\
        & & & MAE coverage & \cellcolor{pale1} 0.2458 & 0.4821 & \cellcolor{pale1} 0.2732 & \cellcolor{pale1} 0.2366 & 0.3440 \\
        \cmidrule(lr){3-4}
        & & \multirow{3}{*}{\parbox{14mm}{\vspace{0mm}\centering After\\fine-tuning}} & MSE & \cellcolor{pale1} 0.0937 & \cellcolor{pale1} 0.0054 & \cellcolor{pale1} 0.0163 & 0.0550 & \cellcolor{pale1} 0.0374 \\
        & & & Variance & \cellcolor{pale1} 0.0021 & \cellcolor{pale1} 0.0002 & \cellcolor{pale1} 0.0008 & \cellcolor{pale1} 0.0035 & \cellcolor{pale1} 0.0027 \\
        & & & MAE coverage & 0.2625 & \cellcolor{pale1} 0.3750 & 0.3468 & 0.2652 & \cellcolor{pale1} 0.3165 \\
        \bottomrule
    \end{tabular}
    }
\end{table}

\subsection{Experiment Results}

As shown in Table \ref{table_results}, we present the prediction performance of the Lag-Llama and Chronos models in both zero-shot and fine-tuned scenarios. We fine-tuned the models using the three in-sample datasets (FRED, World Cup search trends, and EIA Daily) and evaluated them on all five collected datasets.

The table highlights the comparison results in terms of the three metrics between zero-shot and after fine-tuning for all model and dataset combinations. These results indicate a significant improvement in in-sample datasets, with both MSE and variance measures lower than those zero-shot measures. However, improvement in the MAE coverage measures is limited. While the degree of improvement varies across different models, such improvements support the idea that models can be effectively fine-tuned to adapt to time series with distributional shifts. The testing results on the two out-sample datasets (Yahoo and COVID search trends) demonstrate the generalizability of the fine-tuned model on distributional shift data. Although the improvements are more modest compared to the in-sample datasets, they still represent a notable enhancement over the zero-shot performance. This highlights the models' ability to adapt and apply learned patterns to unseen data, thereby validating the robustness of the fine-tuning process with distributional shift data.

The fine-tuning process significantly enhanced the performance of both Lag-Llama and Chronos models. The results demonstrate the utility of the collected datasets in improving model accuracy and consistency, particularly in the presence of distributional shifts.

\section{Discussion}
\label{ch_discussion}

The proposed approach for creating alternative datasets using LLMs and data source APIs demonstrates an advancement in addressing the challenges associated with time series analysis, particularly under data scarcity and distributional shifts. This methodology leverages the comprehensive capabilities of LLMs to identify, retrieve, and augment time series data, ensuring that the collected datasets reflect critical statistical properties essential for robust modeling. This pipeline can be adaptive across various domains, where the availability of comprehensive and high-quality datasets can significantly affect downstream modeling. By leveraging LLMs, we can efficiently explore diverse data sources and identify relevant time series datasets that exhibit the necessary statistical properties. This flexibility makes this approach highly generalizable.

The pipeline's ability to handle distributional shifts is particularly noteworthy. By explicitly targeting datasets that reflect this property, the proposed approach ensures that models trained on these collected datasets are better equipped to handle scenarios where distributional changes occur. Another critical aspect of this approach is the integration of data pruning and augmentation. Data pruning ensures the collected time series samples satisfy the requirements of specific statistical properties. Data augmentation enhances the diversity and quantity of collected datasets, which is beneficial when dealing with limited samples.

One concern is the consistency of the collected data. Variations in data sources, such as differences in sampling intervals and lengths, can introduce noise and biases that may affect the performance of downstream tasks. Ensuring the reliability of the collected data requires validation and quality control measures. Our approach can be adapted to various time resolutions to collect time series data, ensuring that downstream models can analyze datasets with different temporal properties. This adaptability is important in applications like financial markets, where data resolutions can range from milliseconds to days. Furthermore, the potential applications of this pipeline can extend beyond various domains and scenarios, not limited to distributional shift time series datasets.

\section{Conclusion}
\label{ch_conclusion}

This paper presents a novel approach to addressing the challenges of data scarcity and distributional shifts in time series analysis by using LLMs and data source APIs. The proposed pipeline leverages the extensive capabilities of LLMs to explore and identify relevant datasets and collect data samples with distributional shifts. The experiments conducted with time series forecasting foundation models demonstrate the effectiveness of the collected datasets in enhancing model performance and generalization capability. While we focus on time series data with distributional shifts in this work, the generalizability of this approach across various domains and scenarios highlights its potential in data exploration and collection.

\begin{ack}

This paper was prepared for informational purposes in part by the Artificial Intelligence Research group of JPMorgan Chase \& Co and its affiliates (“J.P. Morgan”) and is not a product of the Research Department of J.P. Morgan. J.P. Morgan makes no representation and warranty whatsoever and disclaims all liability for the completeness, accuracy, or reliability of the information contained herein. This document is not intended as investment research or advice, or a recommendation, offer, or solicitation for the purchase or sale of any security, financial instrument, financial product, or service, or to be used in any way for evaluating the merits of participating in any transaction, and shall not constitute a solicitation under any jurisdiction or to any person if such solicitation under such jurisdiction or to such person would be unlawful.

\end{ack}

\AtBeginEnvironment{thebibliography}{\small}
\bibliographystyle{unsrt}

\end{document}